\documentclass[conference,a4paper]{IEEEtran}
\IEEEoverridecommandlockouts

\usepackage[utf8]{inputenc} 
\usepackage[T1]{fontenc}
\usepackage{url}              
\usepackage{cite}             

\usepackage[cmex10]{amsmath}  
\usepackage{mleftright}       
\mleftright                   

\usepackage{graphicx}         
\usepackage{booktabs}         
\usepackage{CustomMacros}
\usepackage{amsfonts, amssymb}
\usepackage{standalone}
\usepackage{cases}

\usepackage{subcaption}
\usepackage{comment}
\usepackage{etoolbox}
\usepackage{float}
\usepackage{algorithm}
\usepackage{algpseudocode}
\usepackage{dsfont}
\usepackage{bbm}
\usepackage{dsfont}

\allowdisplaybreaks

\usepackage{tikz}
\usepackage{tkz-tab}
\usepackage{tikz-3dplot}
\usetikzlibrary{shapes.multipart}
\usetikzlibrary{shapes.symbols}
\usepackage{pgfplots}

\usepackage{etoolbox}
\newcommand{\zerodisplayskips}{%
  \setlength{\abovedisplayskip}{3pt}%
  \setlength{\belowdisplayskip}{3pt}%
  \setlength{\abovedisplayshortskip}{3pt}%
  \setlength{\belowdisplayshortskip}{3pt}}
\appto{\normalsize}{\zerodisplayskips}
\appto{\small}{\zerodisplayskips}
\appto{\footnotesize}{\zerodisplayskips}

\begin{document}

\title{Collaborative Mean Estimation over Intermittently Connected Networks with Peer-To-Peer Privacy
\thanks{This work was supported by the AFOSR award \#002484665, a Huawei Intelligent Spectrum grant, and NSF grants CCF-1908308 \& CNS-2128448.}
}

\author{\IEEEauthorblockN{Rajarshi Saha}
\IEEEauthorblockA{Stanford University}
\and
\IEEEauthorblockN{Mohamed Seif}
\IEEEauthorblockA{Princeton University}
\and
\IEEEauthorblockN{Michal Yemini}
\IEEEauthorblockA{Bar-Ilan University}
\and
\IEEEauthorblockN{Andrea J. Goldsmith}
\IEEEauthorblockA{Princeton University}
\and
\IEEEauthorblockN{H. Vincent Poor}
\IEEEauthorblockA{Princeton University}
}

\maketitle
\begin{abstract}
  This work considers the problem of \textit{Distributed Mean Estimation} (DME) over networks with intermittent connectivity, where the goal is to learn a global statistic over the data samples localized across distributed nodes with the help of a central server. To mitigate the impact of intermittent links, nodes can collaborate with their neighbors to compute local consensus which they forward to the central server. In such a setup, the communications between any pair of nodes must satisfy \textit{local differential privacy} constraints. We study the tradeoff between collaborative relaying and privacy leakage due to the additional data sharing among nodes
  and, subsequently, propose a novel differentially private collaborative algorithm for DME to achieve the optimal tradeoff. Finally, we present numerical simulations to substantiate our theoretical findings.
\end{abstract}

\section{Introduction}
\label{sec:introduction}

Distributed Mean Estimation (DME) is a fundamental statistical problem that arises in several applications, such as model aggregation in federated learning \cite{mcmahan2017communication}, distributed K-means clustering \cite{balcan2013distributed}, distributed power iteration \cite{suresh2017distributed}, etc.
DME presents several practical challenges, which prior research \cite{jhunjhunwala2021leveraging,kar2008distributed,tandon2017gradient, kar_2008_topology_design, kar_2012_nonlinear_observation_model} has considered, including the problem of straggler nodes, where nodes cannot send their data to the parameter server (PS).
Typically, there are two types of stragglers: $\rm (i)$ {\it computation stragglers}, in which nodes cannot finish their local computation within a deadline, and $\rm (ii)$ {\it communication stragglers}, in which nodes cannot transmit their updates due to communication blockage \cite{gapeyenko_mobile_blockers, yasamin, pappas, zavlanos, yemini_et_al:globecom2020, yemini_cloud_cluster}. 
The problem of communication stragglers can be solved by relaying the updates/data to the PS via neighboring nodes. 
This approach was proposed and analyzed in \cite{saha2022colrel,yemini2022_ISIT_robust,yemini2022robust}, where it was shown that the proposed collaborative relaying scheme can be optimized to reduce the expected distance to optimality, both for DME \cite{saha2022colrel} and federated learning \cite{yemini2022_ISIT_robust,yemini2022robust}.

While the works \cite{yemini2022_ISIT_robust,yemini2022robust,saha2022colrel} show that collaborative relaying reduces the expected distance to optimality, exchanging the individual data across nodes incurs privacy leakage caused by the additional estimates that are shared among the nodes.
Nonetheless, this potential breach of privacy has not been addressed in the aforementioned works.
To mitigate the privacy leakage in DME, we require a rigorous privacy notion. 
Within the context of distributed learning, \textit{local differential privacy} (LDP) \cite{dwork2014algorithmic} has been adopted as a gold standard notion of privacy, in which a user can perturb and disclose a sanitized version of its data to an \textit{untrusted} server. LDP ensures that the statistics of the user's output observed by adversaries are indistinguishable regardless of the realization of any input data. In this paper, we focus on the \textit{node-level} LDP where the neighboring nodes, as well as any eavesdropper that can observe the local node-node transmissions during collaborations, cannot infer the realization of the user's data.

There has been extensive research into the design of distributed learning algorithms that are both communication efficient and private (see \cite{kairouz2021advances}  for a comprehensive survey and references therein). It is worth noting that LDP requires a significant amount of perturbation noise to ensure reasonable privacy guarantees. Nonetheless, the amount of perturbation noise can be significantly reduced by considering the intermittent connectivity of nodes in the learning process \cite{balle2018privacy}. The intermittent connectivity in DME amplifies the privacy guarantees; it provides a boosted level of anonymity due to partial communication with the server. Various random node participation schemes have been proposed to further improve the utility-privacy tradeoff in distributed learning, such as Poisson sampling \cite{zhu2019poission}, importance sampling \cite{luo2022tackling, rizk2022federated}, and sampling with/without replacement \cite{balle2018privacy}. In addition, Balle et al. investigated in \cite{balle2020privacy}, the privacy amplification in federated learning via random check-ins and showed that the privacy leakage scales as $O(1/\sqrt{n})$, where $n$ is the number of nodes. In other words, random node participation reduces the amount of noise required to achieve the same levels of privacy that are achieved without sampling.

So far, works in the privacy literature, such as \cite{dwork2014algorithmic,kairouz2021advances,balle2018privacy,zhu2019poission,luo2022tackling,rizk2022federated,balle2020privacy}, have not considered intermittent connectivity along with collaborative relaying, where nodes share their local updates to mitigate the randomness in network connectivity \cite{yemini2022_ISIT_robust,yemini2022robust,saha2022colrel}.
Thus, this paper aims to close this theoretical gap. To this end, we first show that there exists a tradeoff between collaborative relaying and privacy leakage due to data sharing among nodes for DME under intermittent connectivity assumption.
We introduce our system model and proposed algorithm in \S \ref{sec:system_model_for_private_collaboration}, followed by its utility (MSE) and privacy analyses in \S \ref{sec:mean_squared_error_analysis} and \S \ref{sec:privacy_analysis} respectively.
We quantify this tradeoff by formulating it as an optimization problem and solve it approximately due to its non-convexity. Finally, we demonstrate the efficacy of our private collaborative algorithm through numerical simulations.

\section{System Model for Private Collaboration}
\label{sec:system_model_for_private_collaboration}

Consider a distributed system with $n$ nodes, each having a vector $\xv_i \in \Real^d$, $\lVert \xv_i \rVert_2 \leq {\rm R}$ for some known ${\rm R} > 0$. 
The nodes communicate with a parameter server (PS), as well as with each other over intermittent links with the goal of estimating their mean, $\xvo \triangleq \frac{1}{n}\sum_{i=1}^{n}\xv_i$ at the PS (Fig. \ref{fig:communication_model}).
For any estimate $\xvoh$ of the mean, the evaluation metric for any estimate is the mean-square error (MSE), given by $\Ecal \triangleq \mathbb{E}\lVert\xvoh - \xvo\rVert_2^2$.

\subsection{Communication Model}
\label{subsec:communication_model}

\begin{figure}[t]
\centering
\includegraphics[width = 0.8\columnwidth]{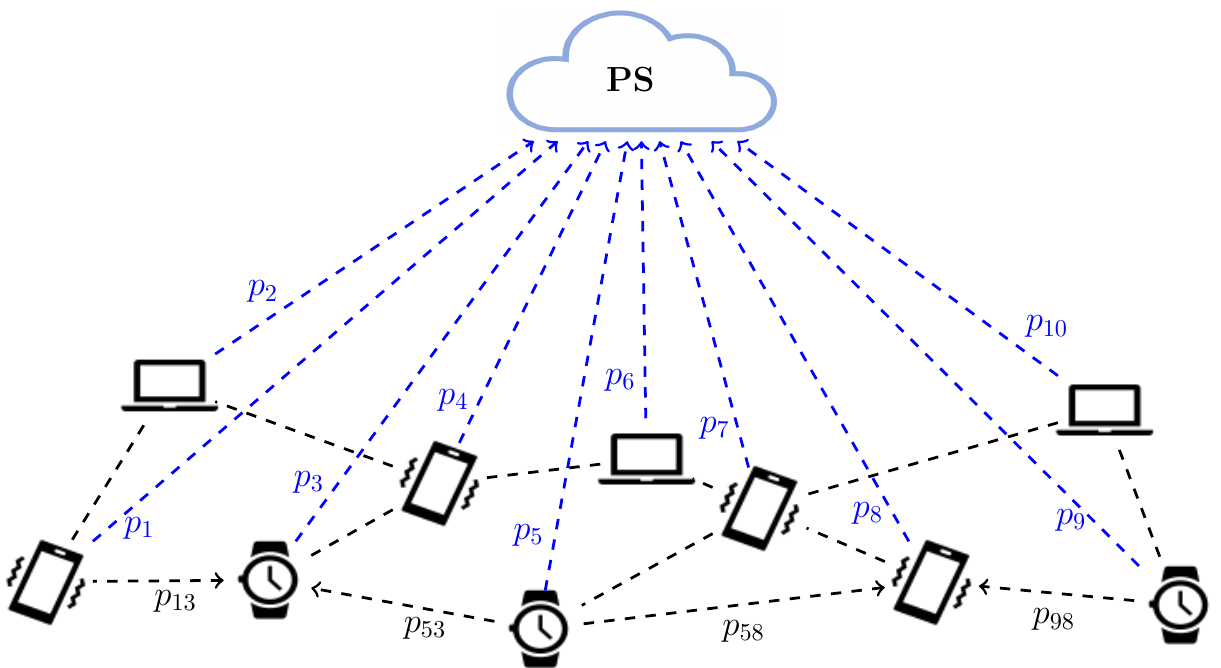}
\caption{ An intermittently connected distributed learning network. Blue and black dotted lines denote intermittent node-PS and node-node connections. Communication between any two nodes must satisfy local differential privacy constraints.}
\vspace{-0.45cm}
\label{fig:communication_model}
\end{figure}

As shown in Fig. \ref{fig:communication_model}, node $i$ can communicate with the PS with a probability $p_i$, with the link modeled using a Bernoulli random variable $\tau_i \sim {\rm Ber}(p_i)$.
Similarly, node $i$ can communicate with another node $j$ with probability $p_{ij}$, i.e., $\tau_{ij} \sim {\rm Ber}(p_{ij})$.
The links between different node pairs are assumed to be statistically independent, i.e., $\tau_i \perp \tau_j$ for $i \neq j$, $\tau_{ij} \perp \tau_{ml}$ for $(i,j) \neq (m,l)$, $(j,i) \neq (m,l)$, and $\tau_{ij} \perp \tau_l$ for $i,j,l \in [n]$.
The correlation due to channel reciprocity between a pair of nodes $i,j \in [n]$ is denoted by ${\rm E}_{\{i,j\}} \equiv \mathbb{E}[\tau_{ij}\tau_{ji}]$.
We assume that $\mathrm{E}_{\{i,j\}}\geq p_{ij}p_{ji}$, i.e., $\mathbb{P}(\tau_{ij}=1|\tau_{ji}=1)\geq \mathbb{P}(\tau_{ij}=1)$
Furthermore, $p_{ii} = 1 \; \forall \; i \in [n]$, and if node $i$ can never transmit to $j$, we set $p_{ij} = 0$. 
We denote $\pv \equiv (p_1, \ldots, p_n)$ and $\Pv \equiv (p_{ij})_{i,j \in [n]} \in [0,1]^{n \times n}$.

\subsection{Privacy Model}
\label{subsec:privacy_model}
The nodes are assumed to be honest but curious. 
They are {\it honest} because they faithfully compute the aggregate of the signals received, however, they are {\it curious} as they might be interested in learning sensitive information about nodes.
Each node uses a local additive noise mechanism to ensure the privacy of its transmissions to neighboring nodes.
We consider local privacy constraints, wherein node $i$ trusts another node $j$ to a certain extent and hence, randomizes its own data accordingly when sharing with node $j$ using a synthetic Gaussian noise (see \cite{{dwork2014algorithmic}}) to respect the privacy constraint while maintaining utility. We present a refresher on differential privacy and Gaussian mechanism in App. \ref{sec:background_on_LDP}.

\subsection{Private Collaborative Relaying for Mean Estimation}
\label{subsec:private_collaborative_relaying_for_mean_estimation}

We now introduce our algorithm, {\sc \bf PriCER}: {\bf Pri}vate {\bf C}ollaborative {\bf E}stimation via {\bf R}elaying.
{\sc PriCER} is a two-stage semi-decentralized algorithm for estimating the mean.
In the first stage, each node $j \in [n]$ sends a scaled and noise-added version of its data to a neighboring node $i \in [n]$ over the intermittent link $\tau_{ji}$.
The transmitted signal is given by,
\begin{equation}
    \xtv_{ji} = \tau_{ji}(\alpha_{ji}\xv_j + \nv_{ji})
\end{equation}
Here, $\alpha_{ji} \geq 0$ is the weight used by node $j$ while sending to node $i$, and $\nv_{ji} \sim \Ncal(\mathbf{0}, \sigma^2\Iv_d)$ is the multivariate Gaussian privacy noise added by node $j$.
Here, $\sigma^2$ is the variance of each coordinate, and $\Iv_d \in \Real^{d \times d}$ is the identity matrix.
We denote the weight matrix by $\Av \equiv (\alpha_{ij})_{i, j \in [n]}$.
Consequently, node $i$ computes the local aggregate of all received signals as,
\begin{equation}
    \xtv_i = \sum_{j \in [n]}\tau_{ji}(\alpha_{ji}\xv_j + \nv_{ji}).
\end{equation}

We quantify our privacy guarantees using the well-established notion of differential privacy \cite{dwork2014algorithmic}.
By observing $\xtv_{ji}$, node $i$ should not be able to distinguish between the events when node $j$ contains the data $\xv_j$ versus when it contains some other data $\xv_j'$.
In other words, we are interested in protecting the local data of node $j$ from a (potentially untrustworthy) neighboring node $i$.
We assume that the privacy noise added by different nodes are uncorrelated, i.e., $\mathbb{E}[\nv_{il}^\top\nv_{jm}] = 0$ for all $i,j,l,m \in [n]$ as long as $i,j,l,m$ are not all equal. 
In the second stage, each node $i$ transmits $\xtv_i$ to the PS over the intermittent link $\tau_i$, and the PS computes the global estimate.
The pseudocode for {\sc PriCER} is given in Algs. \ref{algo:pricer_stage1} and \ref{algo:pricer_stage2}

\begin{algorithm}[t]
\caption{{\sc \bf PriCER-Stage 1} for local aggregation}
\label{algo:pricer_stage1}
{\bf Input}: Non-negative weight matrix $\Av$ 

{\bf Output}: $\widetilde{\xv}_i$ for all $i\in[n]$
\begin{algorithmic}[1]
\For {each $i \in [n]$}
\State Locally generate $\xv_i$
\State Transmit $\xtv_{ij} = \alpha_{ij}\xv_i + \nv_{ij}$ to nodes $j \in [n] : j \neq i$
\State Receive $\xtv_{ji} = \tau_{ji}(\alpha_{ji}\xv_j + \nv_{ji})$ from $j \in [n] : j \neq i$
\State Set $\xtv_{ii} = \alpha_{ii}\xv_i + \nv_{ii}$
\State Locally aggregate available signals: $\xtv_i=\sum_{j\in{n}}\xtv_{ji}$
\State Transmit $\xtv_i$ to the PS
\EndFor
\end{algorithmic}
\end{algorithm}

\begin{algorithm}[t]
\caption{{\sc \bf PriCER-Stage 2} for global aggregation}
\label{algo:pricer_stage2}
{\bf Input}: $\tau_i\xtv_i$ for all $i \in[n]$

{\bf Output}: Estimate of the mean at the PS: $\xvoh$
\begin{algorithmic}[1]
\For {Each $i \in [n]$}
\State Receive $\tau_i\xtv_i$
\EndFor
\State Aggregate the received signals: $\xvoh = \frac{1}{n}\sum_{i \in [n]}\tau_i\xtv_i$
\end{algorithmic}
\end{algorithm}

\section{Mean Squared Error Analysis}
\label{sec:mean_squared_error_analysis}

The goal of {\sc PriCER} is to obtain an unbiased estimate of $\xvo$ at the PS.
Since each node sends its data to all other neighboring nodes, the PS receives multiple copies of the same data.
Lemma \ref{lem:unbiasedness_sufficient_condition}, below gives a sufficient condition to ensure unbiasedness.
This is the same condition as \cite[Lemma 3.1]{saha2022colrel}, and holds true even for {\rm PriCER}.

\begin{lemma}
\label{lem:unbiasedness_sufficient_condition}
Let the weights $\{\alpha_{ij}\}_{i,j \in [n]}$ satisfy
\begin{equation}
\label{eq:unbiasedness_sufficient_condition}
    \sum_{j \in [n]}p_jp_{ij}\alpha_{ij} = 1,
\end{equation}
for every $i \in [n]$. 
Then, $\mathbb{E}\left[{\xvoh \mid \{\xv_i\}_{i \in [n]}}\right] = \xvo = \frac{1}{n}\sum_{i=1}^{n}\xv_i.$
\end{lemma}

We prove this lemma in App. \ref{sec:unbiasedness_sufficient_condition}.

Under this unbiasedness condition, we derive a worst-case upper bound for the MSE in Thm. \ref{thm:mse_PriCER}.

\begin{theorem}
\label{thm:mse_PriCER}
Given $\pv, \Pv$ and $\Av$ such that \eqref{eq:unbiasedness_sufficient_condition} holds, and $\nv_{ij} \sim \Ncal(\mathbf{0}, \sigma^2\Iv_d) \; \forall \; i,j \in [n]$, the MSE with {\rm PriCER} satisfies,
\begin{align}
\label{eq:PriCER_MSE_upper_bound}
    \mathbb{E}\lVert\xvoh - \xv\rVert_2^2 \leq {\rm R}^2\sigma_{\rm tv}^2(\pv,\Pv, \Av) + \sigma_{\rm pr}^2(\pv, \Pv, \sigma),
\end{align}
where $\sigma_{\rm tv}^2(\pv, \Pv, \Av)$ is an upper bound on the variance induced by the stochasticity due to intermittent topology given by,
\begin{align*}
    &\sigma_{\rm tv}^2(\pv, \Pv, \Av) \triangleq \frac{1}{n^2}\left[\sum_{i,j,l\in[n]} p_j(1-p_j)p_{ij}p_{lj}\alpha_{ij}\alpha_{lj} \right. \nonumber\\
    &+\left.\hspace{-2mm}\sum_{i,j\in[n]}\hspace{-1.5mm}p_{ij}p_j(1-p_{ij})\alpha^2_{ij} +\hspace{-2mm}\sum_{i,l\in[n]}\hspace{-1.5mm}p_ip_l(\mathrm{E}_{\{i,l\}}\hspace{-0.5mm}-\hspace{-0.5mm}p_{il}p_{li})\alpha_{li}\alpha_{il}\right],
\end{align*}
and $\sigma_{\rm pr}^2(\pv, \Pv, \sigma)$ is the variance due to the privacy given by,
\begin{align}
    \sigma_{\rm pr}^2(\pv, \Pv, \sigma) \triangleq \frac{1}{n^2}\sum_{i,j \in [n]}p_jp_{ij}\sigma^2d.
\end{align}
\end{theorem}

Thm. \ref{thm:mse_PriCER} is derived in App. \ref{sec:proof_of_theorem_mse_PriCER}.
From \eqref{eq:PriCER_MSE_upper_bound}, we see that $\sigma_{\rm pr}^2(\pv,\Pv, \sigma)$ is the price of privacy.
For a non-private setting, i.e., $\sigma = 0$, the privacy induced variance $\sigma_{\rm pr}^2(\pv, \Pv, \sigma) = 0$, and Thm. \ref{thm:mse_PriCER} simplifies to \cite[Thm. 3.2]{saha2022colrel}.
In the following section, we introduce our privacy guarantee and the corresponding constraints leading to a choice of weight matrix $\Av$ for the optimal Utility (MSE) - Privacy tradeoff of {\sc PriCER}.

\section{Privacy Analysis}
\label{sec:privacy_analysis}

{\sc PriCER} yields privacy guarantees because of two reasons: $\rm (i)$ the local noise added at each node, and $\rm (ii)$ the intermittent nature of the connections.
We consider the local differential privacy when any eavesdropper (possibly including the receiving node) can observe the transmission from node $i$ to node $j$ in stage-$1$ of {\sc PriCER}.
Let us denote the local dataset of node $i$ as ${\cal D}_i$.
In DME, ${\cal D}_i$ is a singleton set and by observing the transmission from node $i$ to node $j$, the eavesdropper should not be able to differentiate between the events $\xv_i \in {\cal D}_i$ and $\xv_i' \in {\cal D}_i$, where $\xv_i' \neq \xv_i$.
The following Thm. \ref{thm:local_privacy_node_node} (derived in App. \ref{sec:proof_local_privacy_node_node}) formally states this guarantee.

\begin{theorem}
\label{thm:local_privacy_node_node}
Given $\nv_{ij}\sim \Ncal(\mathbf{0}, \sigma^2\Iv_d)$, $\xv_i, \xv_i' \in \Real^d$ with $\lVert\xv_i\rVert_2, \lVert\xv_i'\rVert_2 \leq {\rm R}$, and any $\delta_{ij} \in (0,1]$, for pairs $(\epsilon_{ij},p_{ij}\delta_{ij} )$ satisfying
\begin{equation}
\label{eq:eps_expression_node_node_privacy}
    \epsilon_{ij} =
    \begin{cases}
    \Br{2\ln\Paren{\frac{1.25 }{\delta_{ij}}}}^{\frac{1}{2}}\frac{2\alpha_{ij}{\rm R}}{\sigma} \; \; \;  \text{ if } \; p_{ij} > 0, \;\; \text{ and, } \\
    \hspace{15mm}0 \; \;\hspace{13.5mm} \text{ if }\;   p_{ij} = 0,
    \end{cases}
\end{equation}
the transmitted signal from node $i$ to node $j$, $\xtv_{ij} = \tau_{ij}(\alpha_{ij}\xv_i +\nv_{ij})$ is $(\epsilon_{ij}, p_{ij}\delta_{ij})$-differentially private, i.e., it satisfies
\begin{equation}
    \Pr\Paren{\xtv_{ij} \hspace{-0.25mm}\in\hspace{-0.25mm} {\cal S} \hspace{-1mm}\mid\hspace{-1mm} \xv_i \in {\cal D}_i} \leq e^{\epsilon_{ij}}\hspace{-1mm}\Pr\Paren{\xtv_{ij} \hspace{-0.25mm}\in\hspace{-0.25mm} {\cal S} \hspace{-1mm}\mid\hspace{-1mm} \xv_i' \in {\cal D}_i} + p_{ij}\delta_{ij},
\end{equation}
for any measurable set $\cal S$.
\end{theorem}

Setting $\delta := p_{ij}\delta_{ij}$, we can immediately see that intermittent connectivity inherently boosts privacy,
since for the same $\delta$ for any pair $i, j \in [n]$, the privacy level $\epsilon_{ij}$ is proportional to $\ln\left(1.25p_{ij}/\delta\right)^{\frac{1}{2}}$, implying that a smaller $p_{ij}$ leads to a stronger privacy guarantee.
Additionally, from \eqref{eq:eps_expression_node_node_privacy}, the privacy guarantee $\epsilon_{ij}$ is directly related to the weight $\alpha_{ij}$.
That is, if node $i$ trusts node $j$ more, $\epsilon_{ij}$ can be relatively large, and consequently, node $i$ can assign a higher weight to the data it sends to node $j$.
On the other hand, if node $i$ does not trust node $j$ as much, a smaller value will be assigned to $\alpha_{ij}$.
In other words, for the same noise variance $\sigma$, node $i$ will scale the signal $\alpha_{ij}$ so as to reduce the effective signal-to-noise ratio in settings where a higher privacy is required.
Finally,  when $p_{ij} = 0$, {\sc PriCER} ensures $\alpha_{ij} = 0$, implying $\epsilon_{ij} = 0$, i.e., perfect privacy, albeit zero utility.
Our weight optimization (\S \ref{sec:privacy_constrained_weight_optimization}) aims to minimize the MSE subject to the privacy constraints imposed by \eqref{eq:eps_expression_node_node_privacy}.

\section{Privacy Constrained Weight Optimization}
\label{sec:privacy_constrained_weight_optimization}

When deriving the utility-privacy tradeoff, our objective is to minimize the MSE at the PS subject to desired privacy guarantees, namely $(\underline{\epsilon}_{ij},\underline{\delta}_{ij}p_{ij})$ node-node differential privacy.
Here, $\underline{\epsilon}_{ij}, \underline{\delta}_{ij}$ are pre-designated system parameters that quantify the extent to which node $i$ trusts node $j$ (or alternatively, how much it trusts the communication link $i \to j$ against an eavesdropper).
More specifically, we solve:
\begin{align}
    \label{eq:weight_optimization}
    &\min_{\Av,\sigma} \hspace{0.1cm} {\rm R}^2\sigma_{\rm tv}^2(\pv, \Pv, \Av) + \sigma_{\rm pr}^2(\pv,\Pv,\sigma)\nonumber\\
    &\hspace{0.1cm} \text{s.t.: }\hspace{0.1cm} \alpha_{ij}\geq 0,\hspace{0.1cm} \forall \; i,j\in[n],\qquad \hspace{1.5mm}\textrm{(non-negative weights)}\nonumber\\
     &\hspace{0.7cm}  \sum_{j \in [n]}p_jp_{ij}\alpha_{ij} = 1,\hspace{0.1cm} \forall \; i\in[n], \qquad \hspace{2.5mm}\textrm{(unbiasedness)}\nonumber\\
     &\hspace{0.7cm}  \Br{2\ln\Paren{\frac{1.25 }{\underline{\delta}_{ij}}}}^{\frac{1}{2}}\frac{2\alpha_{ij}{\rm R}}{\sigma}\leq \underline{\epsilon}_{ij} \;\; \forall i,j \in [n], \qquad \hspace{-6.5mm}\textrm{(privacy)}\nonumber\\
     &\hspace{0.7cm} \sigma \geq 0. \hspace{5.0cm} \textrm{(privacy)}.
\end{align}

The above optimization \eqref{eq:weight_optimization} is not necessarily convex.
Furthermore, the objective is also not separable with respect to $\Av$ and $\sigma$. Thus, in what follows, we propose an alternate minimization scheme, where we iteratively minimize with respect to $\Av$ and $\sigma$; one variable at a time, keeping the other fixed.
We tie up the components of \S \ref{subsec:optimizing_A_given_sigma} and \S \ref{subsec:optimizing_sigma_given_A} and present the complete {\sc PriCER} weight and variance optimization algorithm in Alg. \ref{algo:opt_alpha_PriCER}. For clarity of presentation, we assume that $p_i>0$ for all $i\in[n]$, so we can have a simple initialization rule.

\subsection{Optimizing variance $\sigma$ for a given weights $\Av$}\label{subsec:optimizing_sigma_given_A}

\begin{figure}[t]
\centering
\includegraphics[width = 0.7\columnwidth]{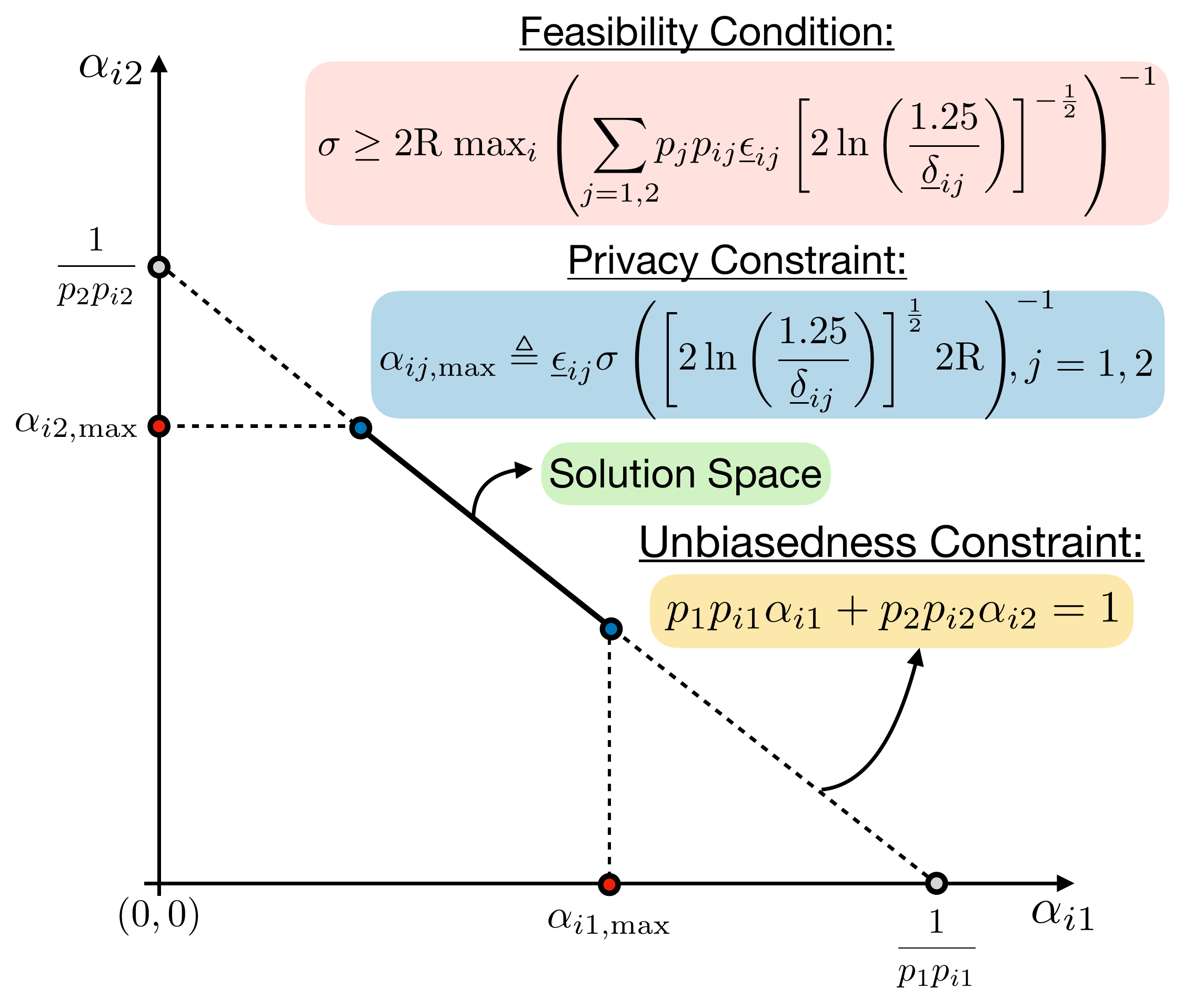}
\vspace{-0.2cm}
\caption{Feasible solution for the optimization problem in \eqref{eq:weight_optimization} with $n = 2$ nodes and $i = 1,2$. Note that $\alpha_{11} = \alpha_{22} = 1$.}
\vspace{-0.6cm}
\label{fig:feasible_set_privacy_constrained_optimization}
\end{figure}

The {\rm non-negative weights}, {\rm unbiasedness}, and {\rm privacy} constraints are present in \eqref{eq:weight_optimization} due to our problem formulation.
However,  when using an alternating optimization we must choose a variance that can fulfill  the unbiasedness condition in the weight optimization stage. 
In other words, {\rm PriCER} needs to add a minimum amount of noise, $\sigma_{\rm thr}$, in order to meet privacy constraints and unbiasedness conditions simultaneously.
Thus, we introduce a necessary condition  to ensure a non-empty feasible set when we optimize the weight for the chosen $\sigma$.

We visualize this in Fig. \ref{fig:feasible_set_privacy_constrained_optimization}.
Note that for a fixed $i \in [n]$, the unbiasedness constraint together with $\alpha_{ij} \geq 0$, defines a hyperplane $\Hcal$ in the positive quadrant of $\Real^n$ with respect to the optimization variables $\{\alpha_{ij}\}_{j \in [n]}$.
Moreover, for $i \in [n]$, the constraints $\alpha_{ij} \geq 0$ and $\alpha_{ij} \leq \underline{\epsilon}_{ij}\sigma\cdot ([2\ln(1.25 /\underline{\delta}_{ij})]^{\frac{1}{2}}2{\rm R})^{-1}$ $ \forall \; j \in [n]$, together define a box $\Bcal$ aligned with the standard basis of $\Real^n$ with one of the vertices at the origin.
The edge length of this box along any of the axes is proportional to $\sigma$
When $\sigma = 0$, i.e., no privacy noise is added, $\Bcal = \mathbf{0}_n$, where $\mathbf{0}_n$ denotes the origin of $\Real^n$.
Since $\Hcal$ does not pass through $\mathbf{0}_n$, \eqref{eq:weight_given_noise_opt_prob_uncorrelated} is infeasible.
This implies there is a minimum value of $\sigma$ so that $\Bcal$ is big enough to have a non-zero intersection with $\Hcal$.
More specifically, we require $\sigma$ such that,
$\sigma\sum_{j \in [n]}p_jp_{ij}\underline{\epsilon}_{ij} \left(\Br{2\ln\Paren{\frac{1.25 }{\underline{\delta}_{ij}}}}^{\frac{1}{2}}2{\rm R}\right)^{-1}\geq 1, \forall\: i\in[n]$,
and hence, we have the last feasibility constraint in \eqref{eq:noise_given_weights_opt_prob_uncorrelated}, where \[\sigma_{\rm{thr}}\triangleq 2{\rm R}\;{\rm max}_{i \in [n]} \left(\sum_{j \in [n]}\hspace{-1mm}p_jp_{ij}\underline{\epsilon}_{ij}[2\ln(\frac{1.25 }{\underline{\delta}_{ij}})]^{-\frac{1}{2}}\right)^{-1}\geq0.\]

We now fix $\Av$ in \eqref{eq:noise_given_weights_opt_prob_uncorrelated} and minimize the {\rm PIV}, i.e.,
\begin{align}
    \label{eq:noise_given_weights_opt_prob_uncorrelated}
    &\min_{\sigma}  \frac{1}{n^2}\sum_{i,j \in [n]}p_jp_{ij}\sigma^2d\nonumber\\
    &\hspace{0.1cm} \text{s.t.: }  \Br{2\ln\Paren{\frac{1.25 }{\underline{\delta}_{ij}}}}^{\frac{1}{2}}\frac{2\alpha_{ij}{\rm R}}{\sigma}\leq \underline{\epsilon}_{ij} \;\; \forall i,j \in [n],\nonumber\\
     &\hspace{0.7cm}\sigma \geq \sigma_{\rm{thr}}.
\end{align}
It can be shown (App. \S \ref{subsec:derivation_optimizing_sigma_given_A}) that the update rule is given by:
\begin{align}
\label{eq:noise_variance_given_A}
    \sigma &= {\rm max}\left\{\max_{i,j\in[n]} \left\{\Br{2\ln\Paren{\frac{1.25 }{\underline{\delta}_{ij}}}}^{\frac{1}{2}}\frac{2\alpha_{ij}{\rm R}}{\underline{\epsilon}_{ij}}\right\},\sigma_{\rm{thr}}\right\}
\end{align}

\subsection{Optimizing weights $\Av$ for a given variance $\sigma$}
\label{subsec:optimizing_A_given_sigma}

Firstly, for a fixed $\sigma$, we minimize the weights $\Av$, i.e.,
\begin{align}
    \label{eq:weight_given_noise_opt_prob_uncorrelated}
    &\min_{\Av} \hspace{0.1cm} {\rm R}^2\sigma_{\rm tv}^2(\pv, \Pv, \Av)\nonumber\\
    &\hspace{0.1cm} \text{s.t.: }\hspace{0.1cm} \alpha_{ij}\geq 0,\hspace{0.1cm} \forall \; i,j\in[n],\quad  \sum_{j \in [n]}p_jp_{ij}\alpha_{ij} = 1,\hspace{0.1cm} \forall \; i\in[n],\nonumber\\
     &\hspace{0.7cm}  \Br{2\ln\Paren{\frac{1.25 }{\underline{\delta}_{ij}}}}^{\frac{1}{2}}\frac{2\alpha_{ij}{\rm R}}{\sigma}\leq \underline{\epsilon}_{ij} \;\; \forall i,j \in [n].
\end{align}

The objective function of problem \eqref{eq:weight_given_noise_opt_prob_uncorrelated} is not convex.
With this in mind, we adopt an approach similar to \cite{saha2022colrel, yemini2022robust} wherein \eqref{eq:weight_given_noise_opt_prob_uncorrelated} is minimized in two iterative stages -- $\rm (i)$ first, a convex relaxation of \eqref{eq:weight_given_noise_opt_prob_uncorrelated} is minimized using Gauss-Seidel method, and $\rm (ii)$ the outcome is subsequently fine-tuned again, using Gauss-Seidel on \eqref{eq:weight_given_noise_opt_prob_uncorrelated}.
The convex relaxation is chosen to be:
\begin{equation}
\label{eq:convex_relaxation_weight_optimization}
    \hspace{-2mm}\min_{\Av} \hspace{0.1cm} {\rm R}^2\overline{\sigma}_{\rm tv}^2(\pv, \Pv, \Av) \; \text{ s.t. the same constraints as } \eqref{eq:weight_given_noise_opt_prob_uncorrelated},
\end{equation}
where the new objective function is,
\begin{align}
    \label{eq:convex_relaxation_objective_function}
    &\overline{\sigma}_{\rm tv}^2(\pv, \Pv, \Av) \triangleq \frac{1}{n^2}\left[\sum_{i,j,l\in[n]} p_j(1-p_j)p_{ij}p_{lj}\alpha_{ij}\alpha_{lj} \nonumber\right.\\
    &\left.+\hspace{-2mm}\sum_{i,j\in[n]}\hspace{-1.5mm}p_{ij}p_j(1-p_{ij})\alpha^2_{ij} +\hspace{-2mm}\sum_{i,l\in[n]}\hspace{-1.5mm}p_ip_l(\mathrm{E}_{\{i,l\}}\hspace{-0.5mm}-\hspace{-0.5mm}p_{il}p_{li})\alpha_{il}^2\right], 
\end{align}

We delegate the complete derivations to App. \S \ref{sec:privacy_constrained_weight_optimization} and only mention the update rules here.
Let us denote the $i^{\rm th}$ row of $\Av$ as $\Av_i$.
Since the objective function of both \eqref{eq:weight_given_noise_opt_prob_uncorrelated} and \eqref{eq:convex_relaxation_objective_function} are separable with respect to $\Av_i$, we can apply Gauss-Seidel iterations on both \eqref{eq:convex_relaxation_weight_optimization} and subsequently on \eqref{eq:weight_given_noise_opt_prob_uncorrelated}.

{\bf Minimizing the convex relaxation \eqref{eq:convex_relaxation_weight_optimization}}:
Let us denote the iterate at the $\ell^{\rm th}$ Gauss-Seidel iteration of the convex relaxation \eqref{eq:convex_relaxation_objective_function}, as $\Av^{(\ell)}$.
Then, the update rule is given by,
\begin{align}
\label{eq:A_hat_iteration_ell_upper_main}
\Av_i^{(\ell)}=\begin{cases}
\widehat{{\Av}}_i^{(\ell)} \;\hspace{2.5mm} \text{ if } i=\ell\;{\rm mod}\;n+n\;\mathds{1}_{\{\ell\;{\rm mod}\;n=0\}},\\
\Av_i^{(\ell-1)} \hspace{0.1mm} \text{ otherwise},
\end{cases}
\end{align}
where $\mathds{1}_{\{\cdot\}}$ denotes the indicator function.
In one iteration, only the $i^{\rm th}$ row, i.e., the weights assigned by the $i^{\rm th}$ node for its neighbors, are updated.
Since Gauss-Seidel performs block-wise descent, the update $\widehat{{\Av}}_i^{(\ell)} \equiv \{\widehat{\alpha}_{ij}\}_{i,j\in[n]}$ can be obtained by formulating the Lagrangian (App. \S \ref{subsec:derivation_optimizing_A_given_sigma}).
Let us denote $\widetilde{w}_{ij}\triangleq\underline{\epsilon}_{ij}\sigma([2\ln(\frac{1.25 }{\underline{\delta}_{ij}})]^{\frac{1}{2}}2{\rm R})^{-1}$ 
and 
\begin{flalign}\label{eq:tilde_alpha_ij}
\overline{\alpha}_{ij}(\lambda_i)\triangleq\left(\frac{-2(1-p_j)\sum_{l\in [n]:l\neq i} p_{l  j}\alpha_{lj}^{(\ell-1)}\hspace{-0.5mm}+\hspace{-0.5mm}\lambda_i }{2[\left(1-p_jp_{ij}\right)+p_i(E_{\{i,j\}}/p_{ij}\hspace{-0.5mm}-\hspace{-0.5mm}p_{ji})]}\right)^{\hspace{-1.5mm}+}.
\end{flalign}

We separate the solution into three scenarios:

\paragraph{\bf{$\boldsymbol{p_i<1}$ and $\boldsymbol{p_jp_{ij}<1}$ for all $\boldsymbol{j\in[n]}$}}
In this case,
\begin{align}
    \label{eq:convex_relaxation_weight_update_1}
    \widehat{\alpha}_{ij}
    = \min\left\{\widetilde{\alpha}_{ij}(\lambda_i),\widetilde{w}_{ij}\right\},
\end{align}
where $\widetilde{\alpha}_{ij}(\lambda_i)$ is given by,
\begin{align}
\label{eq:convex_relaxation_weight_update_2}
\widetilde{\alpha}_{ij}(\lambda_i)
=
\begin{cases}
 \overline{\alpha}_{ij}(\lambda_i) &\text{ if } p_jp_{ij}>0,\\
 0  &\text{ if } p_jp_{ij}=0.
\end{cases}
\end{align}
Here, $(a)^+\triangleq\max\{a,0\}$, and $\lambda_i\geq0$ is set such that $\sum_{j\in[n]}p_jp_{ij}\widehat{\alpha}_{ij}=1$.
$\lambda_i$ is found using bisection search.

\paragraph{\bf{$\boldsymbol{p_i<1}$ and there exists $\boldsymbol{j\neq i}$ such that $\boldsymbol{p_jp_{ij}=1}$}}
Denote $S_i = \sum_{k \in [n]}\mathds{1}_{\{p_kp_{ik} = 1\}}\widetilde{w}_{ik}$. If $S_i\geq 1$, then we choose,
$\widehat{\alpha}_{ij}=\widetilde{w}_{ij}/S_i$ for all $j$ such that $p_jp_{ij}=1$, and $\widehat{\alpha}_{ij}=0$ otherwise.
If $S_i < 1$, we set $\widehat{\alpha}_{ij} = \widetilde{w}_{ij}$ for nodes $j$ that satisfy $p_jp_{ij} = 1$, and subsequently allocate the residual $1-S_i$ of the unbiasedness condition to minimize the objective function.
Similar to \eqref{eq:convex_relaxation_weight_update_1}, this will yield that:
   $ \widehat{\alpha}_{ij}
    = \min\left\{\widetilde{\alpha}_{ij}(\lambda_i),\widetilde{w}_{ij}\right\}$,
%
where $\widetilde{\alpha}_{ij}(\lambda_i)$ is given by,

\begin{align}
\label{eq:convex_relaxation_weight_update_2perfect}
\widetilde{\alpha}_{ij}(\lambda_i)
=
\begin{cases}
 \overline{\alpha}_{ij}(\lambda_i) &\text{ if } p_jp_{ij}\in(0,1),\\
 0  &\text{ if } p_jp_{ij}=0.
\end{cases}
\end{align}

Here, $\lambda_i\geq0$ is such that $\sum_{j: p_jp_{ij} \in (0,1)}p_jp_{ij}\widehat{\alpha}_{ij}=1 - S_i$.

\paragraph{\bf {$\boldsymbol{p_i=1}$}}
In this case, to preserve privacy we set,
\begin{align}\label{eq:convex_relaxation_weight_update_p_i1}
\widehat{\alpha}_{ij}^{(\ell)} =
\begin{cases}
  1 & \text{ if } j=i,\\ 
  0 & \text{ otherwise}.
\end{cases}
\end{align}

{\bf Fine tuning \eqref{eq:weight_given_noise_opt_prob_uncorrelated}}: 
We now fine tune the solution of above by setting it as a warm start initialization and performing Gauss-Seidel on \eqref{eq:weight_given_noise_opt_prob_uncorrelated}.
Then, the update equation for fine tuning is of the same form as \eqref{eq:A_hat_iteration_ell_upper_main} and \eqref{eq:convex_relaxation_weight_update_1}-\eqref{eq:convex_relaxation_weight_update_p_i1}.
However, we plug-in the updated quantity $\widetilde{\alpha}_{ij}(\lambda_i)$ for this case, which is now,
\begin{align}
\label{eq:weight_update_fine_tune}
   \widetilde{\alpha}_{ij}(\lambda_i)
&=\left(\frac{1}{2\left(1-p_jp_{i  j}\right)}\left(-2(1-p_j)\hspace{-2mm}\sum_{l\in [n]:l\neq i} \hspace{-2mm} p_{l  j}\alpha_{lj}^{(\ell-1)}\right.\right. \nonumber \\ &\hspace{5mm}\left.\left.-2p_i(E_{\{i,j\}}/p_{i  j}-p_{j  i})\alpha_{ji}^{(\ell-1)}+\lambda_i\right)\right)^+.
\end{align}

\begin{algorithm}[t]
\caption{MSE-Privacy tradeoff: Joint opt. of $\boldsymbol{A}$ and $\sigma^2$}
\label{algo:opt_alpha_PriCER}
{\bf Input:} Connection probabilities: $\pv>0$, $\Pv$, Pairwise privacy levels: $\{\underline{\epsilon}_{ij}, \underline{\delta}_{ij}\}_{i,j \in [n]}$, Maximal iterations: ${\rm K}$, ${\rm L_1}$, ${\rm L_2}$.\\
{\bf Output}: Weight matrix $\Av^{(\rm K)}$ and privacy noise variance $\sigma^{(\rm K)}$ that approximately solve \eqref{eq:weight_optimization} .

{\bf Initialize}: $\sigma^{(0)}=\sigma_{\rm{thr}}$ and $\Av^{(0)}=\text{diag}\left(\frac{1}{p_{1}},\ldots,\frac{1}{p_{n}}\right)$.

\begin{algorithmic}[1]

\For{$k \gets 0$ \textbf{to} $\rm K-1$}

\State $k \gets k + 1$.

\State Set $\Av^{(k,0)} \gets \Av^{(k-1)}$.

\State Initialize $\ell \gets 0$.

\For{$\ell \gets 0$ \textbf{to} $\rm L_1-1$ {\bf minimize convex relaxation},}

    $\ell \gets \ell + 1$.
    
    $i \leftarrow \ell\mod{n}+n\cdot\mathds{1}_{\{\ell\mod{n}=0\}}$.
    
    Compute $\widehat{\Av}_i^{(k, \ell)}$ according to \eqref{eq:tilde_alpha_ij}-\eqref{eq:convex_relaxation_weight_update_p_i1}.
    
    Set $\Av_i^{(k, \ell)}$ according to \eqref{eq:A_hat_iteration_ell_upper_main}.

\EndFor

\State Warm initialize $\Av^{(k,0)} \gets \Av^{(k,{\rm L})}$, re-initialize $\ell \gets 0$.

\For{$\ell \gets 0$ \textbf{to} $\rm L_2-1$ {\bf fine tune},}

    $\ell \gets \ell + 1$.
    
    $i \leftarrow \ell\mod{n}+n\cdot\mathds{1}_{\{\ell\mod{n}=0\}}$.
    
    Compute $\widehat{\Av}_i^{(k, \ell)}$ according to \eqref{eq:convex_relaxation_weight_update_1}-\eqref{eq:weight_update_fine_tune}.
    
    Set $\Av_i^{(k, \ell)}$ according to \eqref{eq:A_hat_iteration_ell_upper_main}.

\EndFor

\State Set $\Av^{(k)} \gets \Av^{(k, {\rm L})}$.

\State For weights $\Av^{(k)}$, set $\sigma^{(k)}$ according to \eqref{eq:noise_variance_given_A}.
\EndFor
\end{algorithmic}
\end{algorithm}

\section{Numerical Simulations}
\label{sec:numerical_simulations}

\begin{figure}[t]
\centering
  \includegraphics[width=0.77\linewidth,trim={0.3cm 0cm 2.3cm 0cm},clip]{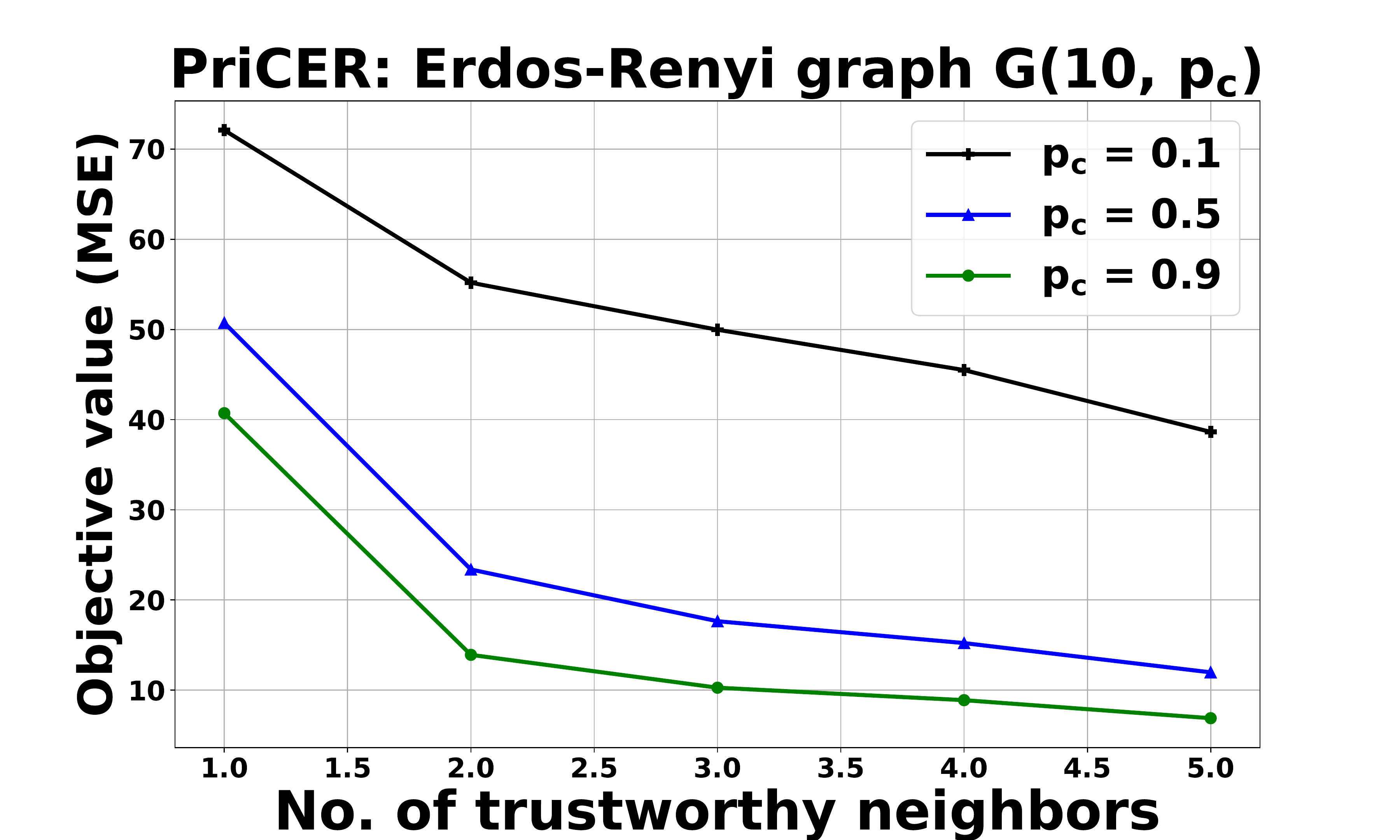}
  \caption{Variation of worst-case MSE with trustworthy neighbors}
  \label{fig:erdos_renyi_vary_trust_ngbrs}
  \vspace{-0.45cm}
\end{figure}

In Fig. \ref{fig:erdos_renyi_vary_trust_ngbrs}, we consider a setup with $n = 10$ nodes that can collaborate over an Erd\H{o}s-R\'enyi topology, i.e., $P_{ij} = {\rm p_c}$ for $j \neq i$ and $P_{ii} = 1$.
The nodes can communicate to the PS with probabilities $\pv = [0.1, 0.1, 0.8, 0.1, 0.1, 0.9, 0.1, 0.1, 0.9, 0.1]$, i.e., only three clients have good connectivity.
Even though any node can communicate with any other node with a non-zero probability, they do not do so because they only trust a small number of immediate neighbors, which is varied along the $\rm x$-axis.
If node $i$ trusts node $j$, we set $\epsilon_{ij} = \epsilon_{\rm high} = 10^3$ (low privacy), otherwise, $\epsilon_{ij} = \epsilon_{\rm low} = 0.1$ (high privacy).
Moreover, $\epsilon_{ii} = \epsilon_{\rm high}$.
We also set $\delta_{ij} = \delta = 10^{-3}$.
$\rm y$-axis shows the (optimized) objective value of \eqref{eq:weight_optimization}, i.e., the (worst-case) upper bound to MSE.
As is evident from Fig. \ref{fig:erdos_renyi_vary_trust_ngbrs}, the  MSE decreases as nodes trust more neighbors, as expected.

\begin{figure}[t]
\centering
  \includegraphics[width=0.77\linewidth,trim={0cm 0cm 2.3cm 0cm},clip]{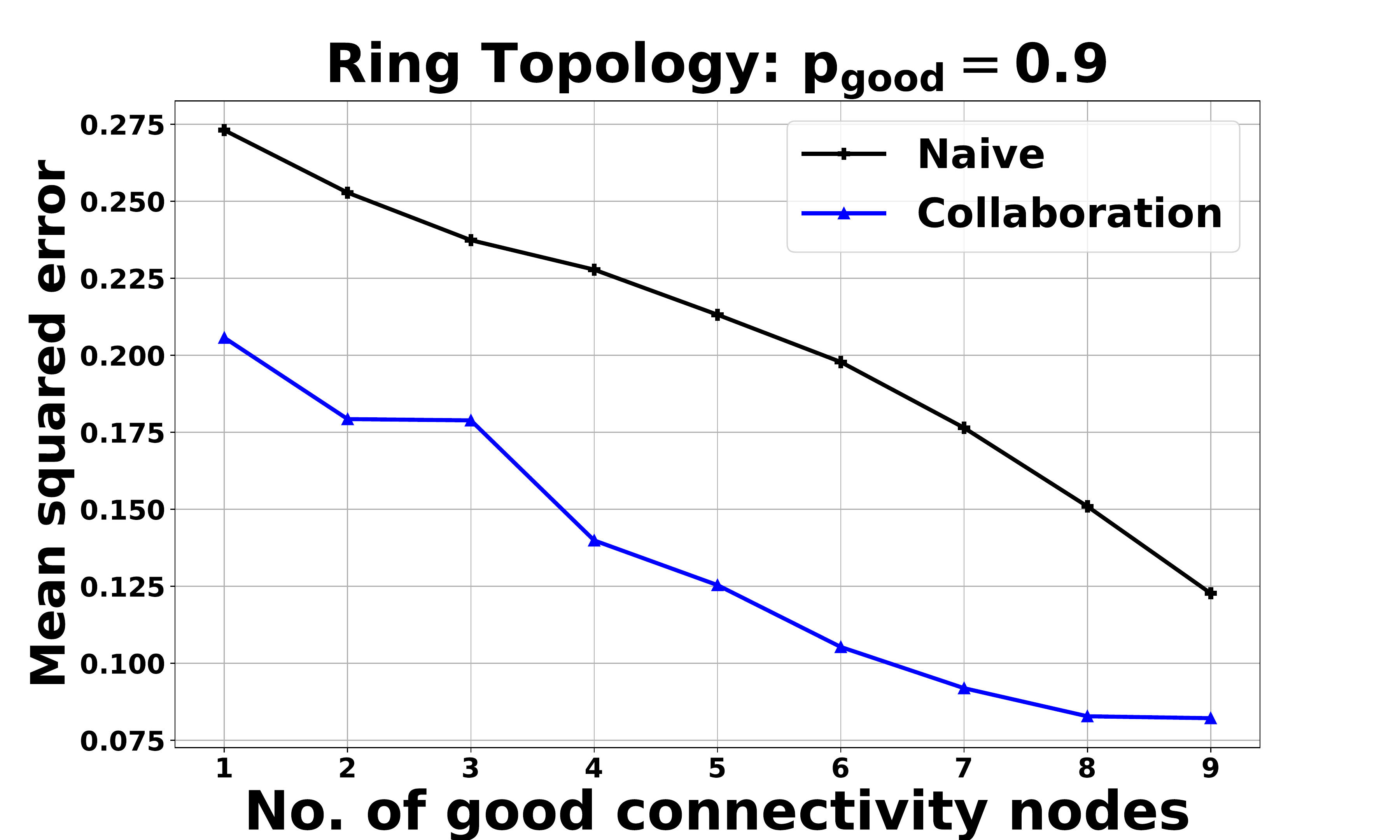}
  \caption{Variation of MSE with number of good connectivity nodes}
  \label{fig:dme_colab_vs_naive}
  \vspace{-0.45cm}
\end{figure}

In Fig. \ref{fig:dme_colab_vs_naive}, we consider that the data at each node is generated from a Gaussian distribution $\Ncal(0,1)$, raised to the power of $3$, and normalized.
This generates a heavy-tailed distribution.
Consequently, if a node that has a vector with a few large coordinate values is unable to convey its data to the PS due to a failed transmission, this can incur a significant MSE.
In this setup only some nodes have good connectivity to the PS, i.e., $p_i = p_{\rm good} = 0.9$, and the remaining have $p_i = p_{\rm bad} = 0.2$
In the na\"ive strategy, the PS averages whatever it successfully receives, i.e., it computes the mean estimate as $\frac{1}{n}\sum_{i \in [n]}\tau_i\xv_i$.
Whereas, in our collaborative strategy, each node trust $6$ other neighbors and can communicate with them with a probability $P_{ij} = 0.8$.
Clearly, {\sc PriCER} achieves a lower MSE than the na\"ive strategy.
The plots are averaged over $50$ realizations.

\vspace{-5px}
\section{Conclusions}
\label{sec:conclusions}
\vspace{-3px}

In this paper, we considered the problem of mean estimation over intermittently connected networks with collaborative relaying subject to peer-to-peer local differential privacy constraints.
The nodes participating in the collaboration do not trust each other completely and, in order to ensure privacy, they scale and perturb their local data when sharing with others.
We have proposed a two-stage consensus algorithm ({\sc PriCER}), that takes into account these peer-to-peer privacy constraints to jointly optimize the scaling weights and noise variance so as to obtain an unbiased estimate of the mean at the PS that minimizes the MSE.
Numerical simulations showed the improvement of our algorithm relative to a non-collaborative strategy in MSE for various network topologies.
Although this work considers peer-to-peer privacy, there can be other sources of privacy leakage such as central DP at the PS too.
Moreover, adding correlated privacy noise may help reduce the MSE even further.
Our future work will include investigating these questions in more detail.

\bibliographystyle{IEEEtran}
\bibliography{references.bib}

\clearpage
\onecolumn
\appendices

\section{A brief refresher on Differential Privacy}
\label{sec:background_on_LDP}

\begin{definition} 
($(\epsilon_{i}, \delta_{i})$-LDP \cite{{dwork2014algorithmic}}) Let $\mathcal{D}_i$ be a set of all possible data points at node $i$. For node $i$, a randomized mechanism $\mathcal{M}_i: \mathcal{D}_{i} \rightarrow \Real^{d}$ is $(\epsilon_{i}, \delta_{i})$ locally differentially private (LDP) if for any $x,x' \in \mathcal{D}_i$, and any measurable subset $\mathcal{S} \subseteq \text{Range}(\mathcal{M}_i)$, we have
\begin{align}
    \operatorname{Pr}(\mathcal{M}_i(x) \in \mathcal{S}) \leq e^{\epsilon_{i}}  \operatorname{Pr}(\mathcal{M}_i(x') \in \mathcal{S}) + \delta_{i}.
\end{align}
The setting when $\delta_{i} = 0$ is referred as pure $\epsilon_{i}$-LDP.\\
\end{definition}

In this work, we analyze the privacy level achieved by {\sc PriCER} algorithm that adds synthetic noise perturbations to privatize its local data. We focus on analyzing the privacy leakage under an additive noise mechanism that is drawn from Gaussian distribution. This well-known perturbation technique is known as Gaussian mechanism, and it provides rigorous privacy guarantees, defined next. 

\begin{definition}(Gaussian Mechanism \cite{dwork2014algorithmic}) \label{defn:Gaussian_mechanism} Suppose a node wants to release a function $f(X)$ of an input $X$ subject to $(\epsilon, \delta)$-LDP. The Gaussian release mechanism is defined as:
\begin{align}
\mathcal{M}(X) \triangleq f(X) + \mathcal{N}(0, \sigma^{2} \mathbf{I}).
\end{align}
If the sensitivity of the function is bounded by $\Delta_f$, i.e., $\| f(x) - f(x')\|_{2}\leq \Delta_f$, $\forall x, x'$, then for any $\delta \in (0,1]$, Gaussian mechanism satisfies $(\epsilon, \delta)$-LDP, where 
\begin{align}
    \epsilon = \frac{\Delta_{f}}{\sigma} \sqrt{2 \log \frac{1.25}{\delta}}. \label{gaussian_noise_add}
\end{align}
\end{definition}

\section{Proof of Lemma \ref{lem:unbiasedness_sufficient_condition}}
\label{sec:unbiasedness_sufficient_condition}

The proof of this is the same as \cite[Lemma 3.1]{saha2022colrel} while taking into account the fact that the privacy noise added is zero-mean Gaussian.
In particular, note that since node $i$ sends $\tau_{ij}(\alpha_{ij}\xv_i + \nv_{ij})$ to node $j$, the total expected contribution of $\xv_i$ at the PS is,
\begin{align}
    &\mathbb{E}\Br{\sum_{j \in [n]}\tau_j\tau_{ij}(\alpha_{ij}\xv_i + \nv_{ij}) \Bigg| \{\xv_i\}_{i \in [n]}} \stackrel{\rm (i)}{=} \mathbb{E}\Br{\sum_{j \in [n]}\tau_j\tau_{ij}\alpha_{ij}\xv_i \Bigg| \{\xv_i\}_{i \in [n]}} + \sum_{j \in [n]}p_jp_{ij}\underbrace{\mathbb{E}[\nv_{ij}]}_{=0},
\end{align}
where $\rm (i)$ follows since $\nv_{ij}$ is independent of $\tau_j$ and $\tau_{ij}$.
Substituting \eqref{eq:unbiasedness_sufficient_condition} completes the proof of the lemma.

\section{Proof of Theorem \ref{thm:mse_PriCER}}
\label{sec:proof_of_theorem_mse_PriCER}

Note that the global estimate at the PS is:
\begin{equation}
\xvoh = \frac{1}{n}\sum_{i \in [n]}\tau_i\sum_{j \in [n]}\tau_{ji}(\alpha_{ji}\xv_j + \nv_{ji}).
\end{equation}
Consequently, the MSE can be written as,
\begin{align}
\label{eq:mse_decomposition_tiv_and_piv}
&\mathbb{E}\left\lVert\frac{1}{n}\sum_{i \in [n]}\tau_i\sum_{j \in [n]}\tau_{ji}(\alpha_{ji}\xv_j + \nv_{ji}) - \frac{1}{n}\sum_{i \in [n]}\xv_i\right\rVert_2^2 \stackrel{\rm (i)}{=} \underbrace{\mathbb{E}\left\lVert \frac{1}{n}\sum_{i \in [n]}\tau_i\sum_{j \in [n]}\tau_{ji}\alpha_{ji}\xv_j - \frac{1}{n}\sum_{i \in [n]}\xv_i  \right\rVert_2^2}_{\text{\rm Topology Induced Variance (TIV)}} +\underbrace{\mathbb{E}\left\lVert \frac{1}{n}\sum_{i \in [n]}\tau_i\sum_{j \in [n]}\tau_{ji}\nv_{ji} \right\rVert_2^2}_{\text{\rm Privacy Induced Variance (PIV)}},
\end{align}
where the expectation is taken over the random connectivity and the local perturbation mechanism. Here, $\rm (i)$ follows because for any $j \in [n]$, the cross term is $\mathbb{E}\left[\sum_{i \in [n]}\tau_j\tau_{ij}\alpha_{ij}\xv_j^{\top}\mathbb{E}[\nv_{ij}]\right] = 0$.

The first term {\rm TIV} in \eqref{eq:mse_decomposition_tiv_and_piv} is solely affected by the intermittent connectivity of nodes.
It can be upper bounded in exactly as is done for ColRel in \cite[Thm. 3.2]{saha2022colrel}, and we get,
\begin{equation}
    {\rm TIV} \leq \frac{{\rm R}^2}{n^2}\;{\rm S}(\pv, \Pv, \Av) := {\rm R}^2\sigma_{\rm tv}^2(\pv, \Pv, \Av).
\end{equation}
To simplify {\rm PIV}, which depends on the privacy noise variance, we push $\tau_j$ inside $\sum_{i \in [n]}$ and interchange $i,j \in [n]$ to get, 
\begin{align}
    \label{eq:PIV_simplification}
    &\frac{1}{n^2}\mathbb{E}\left\lVert \sum_{i,j \in [n]}\tau_j \tau_{ij}\nv_{ij}\right\rVert_2^2 = \frac{1}{n^2}\sum_{i \in [n]}\mathbb{E}\left[\left\lVert \sum_{j \in [n]} \tau_j\tau_{ij}\nv_{ij} \right\rVert_2^2\right] + \frac{1}{n^2}\sum_{\substack{i,l \in [n] \\ i \neq l}}\mathbb{E}\Br{\Paren{\sum_{j \in [n]}\tau_j\tau_{ij}\nv_{ij}}^{\hspace{-2mm}\top}\hspace{-1mm}\Paren{\sum_{m \in [n]}\tau_m\tau_{lm}\nv_{lm}}}.
\end{align}

Expanding the first term of \eqref{eq:PIV_simplification} yields,
\begin{align}
    \underbrace{\frac{1}{n^2}\sum_{i,j \in [n]}p_jp_{ij}\sigma^2d}_{:= \sigma_{\rm pr}^2(\pv, \Pv, \sigma)} + \underbrace{\frac{1}{n^2}\sum_{\substack{i,j,k \in [n] \\ j \neq k}}p_jp_kp_{ij}p_{ik}\mathbb{E}[\nv_{ij}^{\top}\nv_{ik}]}_{ = 0},
\end{align}
where the second term is zero due to our assumption of uncorrelated privacy noise, i.e., $\mathbb{E}[\nv_{ij}^{\top}\nv_{ik}] = 0$.
For the same reason, expanding the second term of \eqref{eq:PIV_simplification} yields,
\begin{align}
    &\frac{1}{n^2}\sum_{\substack{i,l \in [n] \\ i \neq l}}\mathbb{E}\Br{\sum_{j, m \in [n]}\tau_j\tau_{ij}\tau_m\tau_{lm}\nv_{ij}^{\top}\nv_{lm}} = \frac{1}{n^2}\sum_{\substack{i,l \in [n] \\ i \neq l}}\sum_{\substack{j, m \in [n],\\j\neq l, m\neq i}}p_jp_{ij}p_mp_{lm}\underbrace{\mathbb{E}\Br{\nv_{ij}^{\top}\nv_{lm}}}_{=0} +\frac{1}{n^2}\sum_{\substack{i,l \in [n] \\ i \neq l}}p_lp_iE_{\{i,l\}}\underbrace{\mathbb{E}\Br{\nv_{ij}^{\top}\nv_{ji}}}_{=0} = 0.
\end{align}

This completes the proof.

\section{Proof of Theorem \ref{thm:local_privacy_node_node}}
\label{sec:proof_local_privacy_node_node}

We begin by considering the two cases of successful transmission, i.e., $(\tau_{ij} = 1)$ and unsuccessful transmission, i.e., $(\tau_{ij} = 0)$ separately.
Note that when $\tau_{ij} = 0$, we have perfect privacy, i.e.,
\begin{align}
    \label{eq:no_transmission_perfect_privacy}
    &\Pr\Paren{\xtv_{ij} \in {\cal S} \mid \xv_i \in {\cal D}_i, \tau_{ij} = 0} = \Pr\Paren{\xtv_{ij} \in {\cal S} \mid \xv_i' \in {\cal D}_i, \tau_{ij} = 0},
\end{align}
since there is no transmission from node $i$, i.e., $\xtv_{ij} = \mathbf{0}$ irrespective of whether $\xv_i \in {\cal D}_i$ or $\xv_i' \in {\cal D}_i$.
When $\tau_{ij} = 1$, node $j$ receives $\xtv_{ij} = \alpha_{ij}\xv_i + \nv_{ij}$.
The $\ell_2$-sensitivity is,
\begin{equation}
    \label{eq:l2_sensitivity_node_node}
    \sup_{\substack{\xv_i, \xv_i' \text{ s.t. } \\ \lVert \xv_i \rVert_2 \leq {\rm R}, \lVert \xv_i' \rVert_2 \leq {\rm R}}} \lVert \alpha_{ij}\xv_i - \alpha_{ij}\xv_i' \rVert_2 \leq 2\alpha_{ij}{\rm R}.
\end{equation}
Using \eqref{eq:l2_sensitivity_node_node} above, from the guarantees of Gaussian mechanism \cite{dwork2014algorithmic}, when $\tau_{ij} = 1$, we have for any $\delta_{ij} \in (0,1)$, the transmission from node $i$ to node $j$ is $(\epsilon_{ij}, \delta_{ij})$ - differentially private, i.e.,
\begin{align}
    \label{eq:successful_transmission_privacy}
    &\Pr\Paren{\xtv_{ij} \in {\cal S} \mid \xv_i \in {\cal D}_i, \tau_{ij} = 1}\leq e^{\epsilon_{ij}}\Pr\Paren{\xtv_{ij} \in {\cal S} \mid \xv_i' \in {\cal D}_i, \tau_{ij} = 1} + \delta_{ij},
\end{align}
where,
\begin{equation*}
    \epsilon_{ij} = \Br{2\ln\Paren{\frac{1.25}{\delta_{ij}}}}^{\frac{1}{2}}\frac{2\alpha_{ij}{\rm R}}{\sigma}.
\end{equation*}

Combining \eqref{eq:no_transmission_perfect_privacy} and \eqref{eq:successful_transmission_privacy}, we have,
\begin{align}
    &\Pr\Paren{\xtv_{ij} \in {\cal S} \mid \xv_i \in {\cal D}_i} \nonumber \\
    &\qquad= p_{ij}\Pr\Paren{\xtv_{ij} \in {\cal S} \mid \xv_i \in {\cal D}_i, \tau_{ij} = 1}+ (1 - p_{ij})\Pr\Paren{\xtv_{ij} \in {\cal S} \mid \xv_i \in {\cal D}_i, \tau_{ij} = 0} \nonumber \\
    &\qquad= p_{ij}\Paren{e^{\epsilon_{ij}}\Pr\Paren{\xtv_{ij} \in {\cal S} \mid \xv_i' \in {\cal D}_i, \tau_{ij} = 1} + \delta} + (1 - p_{ij})\Pr\Paren{\xtv_{ij} \in {\cal S} \mid \xv_i' \in {\cal D}_i, \tau_{ij} = 0} \nonumber \\
    &\qquad\stackrel{\rm (i)}{\leq} p_{ij}e^{\epsilon_{ij}}\Pr\Paren{\xtv_{ij} \in {\cal S} \mid \xv_i' \in {\cal D}_i, \tau_{ij} = 1} + (1 - p_{ij})e^{\epsilon_{ij}}\Pr\Paren{\xtv_{ij} \in {\cal S} \mid \xv_i' \in {\cal D}_i, \tau_{ij} = 0} + p_{ij}\delta_{ij} \nonumber \\
    &\qquad=e^{\epsilon_{ij}}\Pr\Paren{\xtv_{ij}\in{\cal S} \mid \xv' \in {\cal D}_i} + p_{ij}\delta_{ij}.
\end{align}
Here, $\rm (i)$ holds true since $e^{\epsilon_{ij}} \geq 1$.
Note that when $p_{ij} = 0$, node $i$ can never transmit to node $j$ and as a consequence, we can choose $\alpha_{ij} = 0$.
This completes the proof.

\section{Privacy-constrained weight optimization}
\label{sec:derivation_privacy_constrained_weight_optimization}

\subsection{Optimizing noise variance $\sigma$ for a given $\Av$}
\label{subsec:derivation_optimizing_sigma_given_A}

We start with optimizing $\sigma$ when $\Av$ is fixed, i.e., we solve,
\begin{align}
    \label{eq:noise_given_weights_opt_prob_uncorrelated_restated}
    &\min_{\sigma}  \frac{1}{n^2}\sum_{i,j \in [n]}p_jp_{ij}\sigma^2d\nonumber\\
    &\hspace{0.1cm} \text{s.t.: }  \Br{2\ln\Paren{\frac{1.25 }{\underline{\delta}_{ij}}}}^{\frac{1}{2}}\frac{2\alpha_{ij}{\rm R}}{\sigma}\leq \underline{\epsilon}_{ij} \;\; \forall i,j \in [n],\nonumber\\
     &\hspace{0.7cm}\sigma \geq \sigma_{\rm{thr}}.
\end{align}

It is easy to see that the minimizer of the optimization problem \eqref{eq:noise_given_weights_opt_prob_uncorrelated_restated} above is $\sigma=\max\{\sigma_a,\sigma_{\rm{thr}}\}$, where
\begin{flalign}
\sigma_a &\triangleq \max_{i,j\in[n]} \left\{\Br{2\ln\Paren{\frac{1.25 }{\underline{\delta}_{ij}}}}^{\frac{1}{2}}\frac{2\alpha_{ij}{\rm R}}{\underline{\epsilon}_{ij}}\right\}.
\end{flalign} 

\subsection{Optimizing weight matrix $\Av$ for a given $\sigma$}
\label{subsec:derivation_optimizing_A_given_sigma}

We now look how to optimize for $\Av$ when the noise variance $\sigma$ is fixed.
As described in \S \ref{sec:privacy_constrained_weight_optimization}, we first minimize a convex relaxation, and then subsequently fine tune the solution.
We look at the finer details now.

{\bf Solving the convex relaxation}: From \cite[Lemmas $2$ and $7$]{yemini2022robust}, it can be seen, $\rm (i)$ $\sigma_{\rm tv}^2(\pv, \Pv, \Av) \leq \overline{\sigma}_{\rm tv}^2(\pv, \Pv, \Av)$, and $\rm (ii)$ $\overline{\sigma}_{\rm tv}^2(\pv, \Pv, \Av)$ is convex in $\Av$, showing \eqref{eq:convex_relaxation_weight_optimization} (restated below as \eqref{eq:convex_relaxation_problem_restated}) is a convex relaxation of \eqref{eq:weight_given_noise_opt_prob_uncorrelated}.
\begin{align}
    \label{eq:convex_relaxation_problem_restated}
    &\min_{\Av} \hspace{0.1cm} {\rm R}^2\overline{\sigma}_{\rm tv}^2(\pv, \Pv, \Av)\nonumber\\
    &\hspace{0.1cm} \text{s.t.: }\hspace{0.1cm} \alpha_{ij}\geq 0,\hspace{0.1cm} \forall \; i,j\in[n],\nonumber\\
     &\hspace{0.7cm}  \sum_{j \in [n]}p_jp_{ij}\alpha_{ij} = 1,\hspace{0.1cm} \forall \; i\in[n],\nonumber\\
     &\hspace{0.7cm} \alpha_{ij}\leq \widetilde{w}_{ij}, \;\; \forall i,j \in [n].
\end{align}

Let $\Av_i$ denotes the $i^{\rm th}$ row of $\Av$.
As the domain of \eqref{eq:convex_relaxation_problem_restated} is separable with respect to $\Av_i$, we use Gauss-Seidel method \cite[Prop. 2.7.1]{Bertsekas_nonlinear_programming} to converge to the minimizer of \eqref{eq:convex_relaxation_problem_restated}.
Subsequently, we use the solution obtained from above as the starting point to perform Gauss-Seidel iterations again on the original problem \eqref{eq:weight_given_noise_opt_prob_uncorrelated} so as to converge to a stationary point.

Let  $\Av_i^{(\ell)}$ denote the iterate in the $\ell^{\rm th}$ iteration of the Gauss-Seidel method of \eqref{eq:convex_relaxation_problem_restated}.
Here, $\Av_{i}$ denotes the minimize of \eqref{eq:convex_relaxation_objective_function}.
Denote the initial iteration by $\Av^{(0)}$. 
We can improve our iterates using Gauss-Seidel method until convergence to an optimal point of \eqref{eq:convex_relaxation_objective_function} as follows. 
At every iteration $\ell$ we compute $\Av^{(\ell)}$ using,
\begin{equation}
\label{eq:A_hat_iteration_ell_upper}
\Av_i^{(\ell)}=\begin{cases}
\widehat{{\Av}}_i^{(\ell)} \;\hspace{6mm} \text{ if } \ell\;{\rm mod}\;n+n\cdot\mathds{1}_{\{\ell\;{\rm mod}\;n=0\}}=i,\\
\Av_i^{(\ell-1)} \hspace{3.5mm} \text{ otherwise},
\end{cases}
\end{equation}
where,
\begin{align}
\label{eq:opt_alpha_allocation_iterative_upper}
\widehat{\Av}_{i}^{(\ell)}&=\arg\min\left[\sum_{j\in[n]}p_jp_{i  j}\left(1-p_jp_{ij}\right)\alpha_{ij}^2+\;2\sum_{l\in[n],l\neq i}\sum_{j\in[n]} p_j(1-p_j)p_{ij}p_{l  j}\alpha_{ij}\alpha_{lj}^{(\ell-1)}+\sum_{j\in[n]}p_ip_j(E_{\{i,j\}}-p_{i  j}p_{j  i})\alpha^2_{ij}\right],\nonumber\\
&\quad\text{s.t.:}\sum_{j \in [n]}p_jp_{ij}\alpha_{ij}=1, \;\;  \alpha_{ij}\geq 0,\quad \forall j\in[n],\nonumber\\
&\quad \alpha_{ij}\leq \underline{\epsilon}_{ij}\sigma\left(\Br{2\ln\Paren{\frac{1.25 }{\underline{\delta}_{ij}}}}^{\frac{1}{2}}2{\rm R}\right)^{-1}\hspace{-2mm}, \forall i,j \in [n].
\end{align}

The Lagrangian of \eqref{eq:opt_alpha_allocation_iterative_upper} is
\begin{flalign*}
L(\Av_{i}^{(\ell)},\lambda_i) &= \sum_{j\in[n]}p_jp_{i  j}\left(1-p_jp_{i  j}\right)\alpha_{ij}^2 +2\sum_{l\in[n],l\neq i}\sum_{j\in[n]} p_j(1-p_j)p_{i  j}p_{l  j}\alpha_{ij}\alpha_{lj}^{(\ell-1)}+\sum_{j\in[n]}p_ip_j(E_{\{i,j\}}-p_{i  j}p_{j  i})\alpha^2_{ij}\nonumber\\
&-\lambda_i\left(\sum_{j\in[n]} p_jp_{i  j}\alpha_{ij}-1\right)-\mu_{ij}\alpha_{ij}+\nu_i\left(\alpha_{ij}- \underline{\epsilon}_{ij}\sigma\cdot \left(\Br{2\ln\Paren{\frac{1.25 }{\underline{\delta}_{ij}}}}^{\frac{1}{2}}2{\rm R}\right)^{-1}\right).
\end{flalign*}

Evaluating the gradients, 
\begin{flalign*}
\frac{\partial L(\Av_{i}^{(\ell)},\lambda_i)}{\partial \alpha_{ij}}&=
2p_j[p_{ij}\left(1-p_jp_{ij}\right)+p_i(E_{\{i,j\}}-p_{i  j}p_{j  i})]\alpha_{ij}+2p_j(1-p_j)p_{ij}\sum_{l\in [n]:l\neq i} p_{  lj}\alpha_{lj}^{(\ell-1)}\nonumber\\
&\quad-\lambda_ip_{ij} p_j-\mu_{ij}+\nu_{ij},\\ 
\frac{\partial L(\Av_{i}^{(\ell)},\lambda_i)}{\partial \lambda_i} &= 1-\sum_{j\in[n]} p_jp_{ij}\alpha_{ij},\nonumber\\ 
\frac{\partial L(\Av_{i}^{(\ell)},\lambda_i)}{\partial \mu_{ij}} &= -\alpha_{ij},\nonumber\\ 
 \frac{\partial L(\Av_{i}^{(\ell)},\lambda_i)}{\partial \nu_{ij}} &= \alpha_{ij}-\widetilde{w}_{ij}.
\end{flalign*}
From the KKT conditions, we can optimize $\widehat{\alpha}_{ij}^{(\ell)}$ according to \eqref{eq:tilde_alpha_ij}-\eqref{eq:convex_relaxation_weight_update_p_i1}.
When the Lagrange multiplier $\lambda_i\geq0$ is used, we set it such that $\sum_{j\in[n]}p_jp_{ij}\alpha_{ij}(\lambda_i)=1$.
We can find  $\lambda_i$ using the bisection method over the interval: 
\begin{align}
    &\left[0,\max_{j:p_jp_{i  j\in(0,1)}}\left\{\frac{2[\left(1-p_jp_{ij}\right)+p_i({\rm E}_{\{i,j\}}/p_{ij}-p_{ji})]}{p_jp_{ij}}+\;2(1-p_j)\sum_{l\in [n]:l\neq i} p_{l  j}\alpha_{lj}^{(\ell-1)}\right\}\right].
\end{align}

{\bf Fine tuning the solution}:
We are now ready to fine tune the solution of the convex relaxation obtained from above.
The derivation of the update equations for this is the similar to that of the convex relaxation.
We skip the details and only point out the differences.
The original problem \eqref{eq:weight_given_noise_opt_prob_uncorrelated} and \eqref{eq:convex_relaxation_problem_restated} differ in the last term, i.e., \eqref{eq:weight_given_noise_opt_prob_uncorrelated} has $\alpha_{ij}\alpha_{ji}$ instead of $\alpha_{ij}^2$.
Denoting the $\ell^{\rm th}$ iterate of the (fine-tuning) Gauss-Seidel iterations as $\Av_i$, the update equation is the same as \eqref{eq:A_hat_iteration_ell_upper}, with $\Av_i^{\ell}$ instead of $\Av_i^{\ell}$.
Consequently, the corresponding expressions for \eqref{eq:opt_alpha_allocation_iterative_upper}, and the new Lagrangian have $2\sum_{j\in[n]}p_ip_j(E_{\{i,j\}}-p_{ij}p_{ji})\alpha_{ij}\alpha_{ji}$, instead of $\sum_{j\in[n]}p_ip_j(E_{\{i,j\}}-p_{ij}p_{ji})\alpha^2_{ij}$.
Once again, using the KKT conditions, we can optimize $\widehat{\alpha}_{ij}^{(\ell)}$ according to \eqref{eq:convex_relaxation_weight_update_1}-\eqref{eq:convex_relaxation_weight_update_p_i1},
%
where $\widetilde{\alpha}_{ij}(\lambda_i)$ is now given by,
\begin{align}
\label{eq:weight_update_fine_tune_restated}
    \widetilde{\alpha}_{ij}^{(\ell)}(\lambda_i)
&=\left(\frac{1}{2\left(1-p_jp_{i  j}\right)}\left(-2(1-p_j)\hspace{-2mm}\sum_{l\in [n]:l\neq i} \hspace{-2mm} p_{l  j}\alpha_{lj}^{(\ell-1)}-2p_i(E_{\{i,j\}}/p_{i  j}-p_{j  i})\alpha_{ji}^{(\ell-1)}+\lambda_i\right)\right)^+.
\end{align}
As before, $\lambda_i\geq0$ is set such that $\sum_{j\in[n]}p_jp_{ij}\alpha_{ij}(\lambda_i)=1$, and once again, we can find $\lambda_i$ using the bisection method over the interval:
\begin{align}
    &\left[0,\max_{j:p_jp_{i  j\in(0,1)}}\left\{\frac{2\left(1-p_jp_{i  j}\right)}{p_jp_{ij}}+2(1-p_j)\sum_{l\in [n]:l\neq i} p_{l  j}\alpha_{lj}^{(\ell-1)}+\;2p_i\left(\frac{{\rm E}_{\{i,j\}}}{p_{ij}}-p_{j  i}\right)\alpha_{ji}^{(\ell-1)}\right\}\right].
\end{align}

This completes the derivation in this section.

\end{document}